# Photon wave function and *zitterbewegung*

**Zhi-Yong Wang, Cai-Dong Xiong, Qi Qiu**

*School of Optoelectronic Information, University of Electronic Science and Technology of China, Chengdu 610054, CHINA*

In terms of a photon wave function corresponding to the (1, 0)+(0, 1) representation of the Lorentz group, the radiation and Coulomb fields within a source-free region can be described unitedly by a Lorentz-covariant Dirac-like equation. In our formalism, the relation between the positive- and negative-energy solutions of the Dirac-like equation corresponds to the duality between the electric and magnetic fields, rather than to the usual particle-antiparticle symmetry. The *zitterbewegung* (ZB) of photons is studied via the momentum vector of the electromagnetic field, which shows that only in the presence of virtual longitudinal and scalar photons, the ZB motion of photons can occur, and its vector property is described by the polarization vectors of the electromagnetic field.

## I. INTRODUCTION

In spite of the well-known conclusion that photons cannot be spatially localized [1, 2], there have been many attempts to introduce the concept of photon wave function [3-16], and some recent experimental and theoretical studies have shown that photons can be localized in space [17-19]. In fact, in relativistic quantum theory, one can extend the concept of wave function interpreted as a probability (-density) amplitude to those of which interpreted as a charge-density amplitude, energy-density amplitude or particle-number density amplitudes, etc. All these imply that there remains a room for us to introduce the concept of photon wave function.

On the other hand, an old interest in the *zitterbewegung* (ZB) of the Dirac electron has recently been rekindled by the investigations on spintronics, graphene, superconducting systems, etc. [20-26]. In particular, the first proposal for simulating the massless and massive Dirac equation in a quantum optical system has been presented in Ref. [27], and there are recently important progresses in improving the predictions for detecting ZB and relating them to Schrödinger cats in trapped ions [28, 29]. Contrary to these, there has not achieved any progress in the investigation on the ZB of photons since some preliminary studies have been presented [30].

In this paper, a photon wave function interpreted as the energy-density amplitude of the electromagnetic field is introduced, its equation of motion (i.e., the Dirac-like equation) is studied at both classic and quantum level, and some symmetries of our theory are discussed from a new point of view. The ZB of photons is investigated, which shows that both our formalism and the traditional one present the same conclusion. Our work is different from the previous literatures mainly in the following aspects:

(1) Our photon wave function is different from those presented in the previous literatures. For example, in Ref. [7], the six-component photon wave function is actually a counterpart of the Weyl bispinor and corresponds to the chiral representation, while our photon wave function should be regarded as a counterpart of the Dirac bispinor and corresponds to the $(1,0)\oplus(0,1)$ representation of the Lorentz group (the symbol $\oplus$ denotes the direct sum);

(2) In the previous literatures, the negative-energy solutions are always thrown away by hand, which is in conflict with quantum electrodynamics (QED); in this paper the negative-energy solutions are reasonably contained;

(3) In this paper, we clarify the relation between the four kinds of photons and the three

polarization vectors contained in photon wave function. To present a quantization theory based on our formalism, all of our commutation relations are derived from QED rather than introduced by hand;

(4) Our investigation on the ZB of photons is very different from those in the previous literatures (see for example, Ref. [30]). For example, we study the ZB of photons via the momentum of the electromagnetic field so as to avoid introducing a controversial position operator for photons, and we show the ZB of the photon cannot occur in the absence of virtual longitudinal and scalar photons.

In the following we apply the natural units of measurement ($\hbar = c = 1$) and the four-dimensional (4D) space-time metric tensor is taken as $g_{\mu\nu} = \text{diag}(1,-1,-1,-1)$ ($\mu,\nu = 0,1,2,3$). Repeated indices must be summed according to the Einstein rule.

## II. PHOTON WAVE FUNCTION AND THE DIRAC-LIKE EQUATION

As we know, in QED the electromagnetic field is usually described by a 4D electromagnetic potential that transforms according to the (1/2, 1/2) representation of the Lorentz group. However, the source-free electromagnetic field can be described directly via electromagnetic field intensities, such that it can be described in a form closely parallels to the Dirac theory of electron. In vacuum and outside a source, the electric field intensity, $\boldsymbol{E} = (E_1, E_2, E_3)$, and the magnetic field intensity, $\boldsymbol{B} = (B_1, B_2, B_3)$, satisfy the Maxwell equations ($\hbar = c = 1$)

$$\nabla \times \boldsymbol{E} = -\partial_t \boldsymbol{B}, \quad \nabla \times \boldsymbol{B} = \partial_t \boldsymbol{E}, \tag{1}$$

$$\nabla \cdot \boldsymbol{E} = 0, \quad \nabla \cdot \boldsymbol{B} = 0. \tag{2}$$

In the present case Eq. (2) is actually contained by Eq. (1) and then the Maxwell equations can be represented solely by Eq. (1). The vectors $\boldsymbol{E}$ and $\boldsymbol{B}$ can also be expressed as the column matrices: $\boldsymbol{E}=(E_1\ \ E_2\ \ E_3)^{\mathrm{T}}$ and $\boldsymbol{B}=(B_1\ \ B_2\ \ B_3)^{\mathrm{T}}$ (the superscript T denotes the matrix transpose), by them a photon wave function can be introduced as follows:

$$\psi=\frac{1}{\sqrt{2}}\begin{pmatrix}\boldsymbol{E}\\ i\boldsymbol{B}\end{pmatrix}. \tag{3}$$

Moreover, by means of the 3×3 unit matrix $I_{3\times3}$ and the matrix vector $\boldsymbol{\tau}=(\tau_1,\tau_2,\tau_3)$ with the matrix components

$$\tau_1=\begin{pmatrix}0&0&0\\0&0&-i\\0&i&0\end{pmatrix},\ \tau_2=\begin{pmatrix}0&0&i\\0&0&0\\-i&0&0\end{pmatrix},\ \tau_3=\begin{pmatrix}0&-i&0\\i&0&0\\0&0&0\end{pmatrix}, \tag{4}$$

we define a 4D matrix vector $\beta^{\mu}=(\beta^0,\boldsymbol{\beta})=(\beta^0,\beta^1,\beta^2,\beta^3)$ ($\mu=0,1,2,3$), where

$$\beta^0=\beta_0\equiv\begin{pmatrix}I_{3\times3}&0\\0&-I_{3\times3}\end{pmatrix},\ \beta^i=-\beta_i\equiv\begin{pmatrix}0&\tau_i\\-\tau_i&0\end{pmatrix},\ i=1,2,3. \tag{5}$$

Using Eqs. (3)-(5) one can rewrite the Maxwell equations as a *Dirac-like equation*

$$i\beta^{\mu}\partial_{\mu}\psi(x)=0\text{, or }i\partial_t\psi(x)=\hat{H}\psi(x), \tag{6}$$

where $\hat{H}=-i\beta_0\boldsymbol{\beta}\cdot\nabla$ represents the Hamiltonian of photons outside a source. Similar to the Dirac theory of electron, let $\hat{\boldsymbol{L}}=\boldsymbol{x}\times(-i\nabla)$ be the orbital angular momentum operator, one can easily prove $[\hat{H},\hat{\boldsymbol{L}}+\boldsymbol{S}]=0$, where $\boldsymbol{S}=I_{2\times2}\otimes\boldsymbol{\tau}$ represents the spin matrix of photons (the symbol $\otimes$ stands for the direct product), it implies that the spin of photons is one because of $\boldsymbol{S}\cdot\boldsymbol{S}=1(1+1)I_{6\times6}$. Moreover, let $\psi^{+}$ denote the Hermitian adjoint of $\psi$ and $\bar{\psi}\equiv\psi^{+}\beta^0$, the quantity $\bar{\psi}(x)\psi(x)$ [$=(\boldsymbol{E}^2-\boldsymbol{B}^2)/2$] is a Lorentz scalar, while $\psi^{+}(x)\psi(x)$ [$=(\boldsymbol{E}^2+\boldsymbol{B}^2)/2$] corresponds to the energy density of the electromagnetic field. Therefore, as our photon wave function, the 6×1 spinor $\psi(x)$ should be interpreted as the energy-density amplitude of the electromagnetic field, it transforms according to the $(1,0)\oplus(0,1)$ representation of the Lorentz group such that the Dirac-like equation (6) has a Lorentz-covariant form (see Appendix A). There exists a close analogy between our theory and the Dirac theory of electron (see Appendix B).

One can prove that $(\beta^{\mu}\partial_{\mu})(\beta_{\nu}\partial^{\nu})=\partial^{\mu}\partial_{\mu}+\Omega$ with $\Omega\psi(x)=0$ being identical with the transverse conditions [given by Eq. (2)], such that using Eq. (6) one can obtain the wave equation $\partial^{\mu}\partial_{\mu}\psi(x)=0$, where ($\nabla_i=\partial/\partial x^i$, $i=1,2,3$)

$$\Omega=I_{2\times 2}\otimes\begin{pmatrix}\nabla_1\nabla_1 & \nabla_1\nabla_2 & \nabla_1\nabla_3\\ \nabla_2\nabla_1 & \nabla_2\nabla_2 & \nabla_2\nabla_3\\ \nabla_3\nabla_1 & \nabla_3\nabla_2 & \nabla_3\nabla_3\end{pmatrix}. \tag{7}$$

The Dirac-like equation (6) is valid for all kinds of electromagnetic fields outside a source.

## III. PHOTON WAVE FUNCTION AS THE GENERAL SOLUTION OF THE DIRAC-LIKE EQUATION

Let $k\cdot x\equiv k_{\mu}x^{\mu}=\omega t-\boldsymbol{k}\cdot\boldsymbol{x}$, where $\omega$ is the frequency and $\boldsymbol{k}$ the wave-number vector of photons, and in the natural units of measurement ($\hbar=c=1$), $k^{\mu}=(\omega,\boldsymbol{k})$ also denotes the 4D momentum of single photons. Substituting $\varphi(k)\exp(-\mathrm{i}k\cdot x)$ into Eq. (6) one has $\det(\omega-\boldsymbol{\chi}\cdot\boldsymbol{k})=0$, using it and Eq. (2) one has $\omega_{\lambda}=|\boldsymbol{k}_{\lambda}|$, $\lambda=\pm1,0$, and

$$k_{\lambda}^{\mu}=\begin{cases}k_{\pm1}^{\mu}=(\omega_{\pm1},\boldsymbol{k}_{\pm1})=(\omega,\boldsymbol{k}),\ \lambda=\pm1\\ k_{0}^{\mu}=(\omega_{0},\boldsymbol{k}_{0})=(0,0,0,0),\ \lambda=0\end{cases}. \tag{8}$$

As shown later, the $\lambda=\pm1$ solutions describe the transverse photons, while the $\lambda=0$ solution describes the admixture of the longitudinal and scalar photons. Eq. (8) shows that $k_{\pm1}^{\mu}=(\omega_{\pm1},\boldsymbol{k}_{\pm1})$ denotes the 4D momentum of the transverse photons, while $k_{0}^{\mu}=(\omega_{0},\boldsymbol{k}_{0})$ is the total 4D momentum of the longitudinal and scalar photons, which is in agreement with the traditional conclusion that the contributions of the longitudinal and scalar photons to the energy and momentum of the electromagnetic field cancel each other. For completeness, the $\lambda=0$ solution can formally be remained in the general solution, and its contributions to the final results vanish by taking the limit of $k_0^\mu \to 0$ (or equivalently, by letting the average of the creation or annihilation operator of the $\lambda = 0$ photons vanish).

Under the transformation of $\boldsymbol{E} \leftrightarrow i\boldsymbol{B}$, the Maxwell equations (1) and (2) are invariant, that is, there exists a duality between the electric and magnetic fields. As a result, the Dirac-like equation (6) has two alternative sets of fundamental solutions (they separately form an orthonormal and complete set):

$$\phi_{\pm k,\lambda}(x) = (\omega/V)^{1/2} f(k,\lambda)\exp(\mp ik\cdot x), \quad \lambda = \pm 1, 0, \tag{9a}$$

or

$$\phi'_{\pm k,\lambda}(x) = (\omega/V)^{1/2} g(k,\lambda)\exp(\mp ik\cdot x), \quad \lambda = \pm 1, 0, \tag{9b}$$

where $V = \int d^3x$ represents the normalization volume (the box normalization is adopted such that all momentum integrals are expressed as discrete momentum sums), and

$$\begin{cases} f(k,\lambda) = \dfrac{1}{\sqrt{1+\lambda^2}}\begin{pmatrix} \varepsilon(\boldsymbol{k},\lambda) \\ \lambda\varepsilon(\boldsymbol{k},\lambda) \end{pmatrix} \\ g(k,\lambda) = \dfrac{1}{\sqrt{1+\lambda^2}}\begin{pmatrix} \lambda\varepsilon(\boldsymbol{k},\lambda) \\ \varepsilon(\boldsymbol{k},\lambda) \end{pmatrix} \end{cases}, \quad \lambda = \pm 1, 0, \tag{10}$$

where [$\varepsilon^*(\boldsymbol{k},-1)$ is the complex conjugate of $\varepsilon(\boldsymbol{k},-1)$, and so on]

$$\varepsilon(\boldsymbol{k},1) = \varepsilon^*(\boldsymbol{k},-1) = \frac{1}{\sqrt{2}|\boldsymbol{k}|}\begin{pmatrix} \dfrac{k_1k_3 - ik_2|\boldsymbol{k}|}{k_1 - ik_2} \\ \dfrac{k_2k_3 + ik_1|\boldsymbol{k}|}{k_1 - ik_2} \\ -(k_1 + ik_2) \end{pmatrix}, \quad \varepsilon(\boldsymbol{k},0) = \frac{1}{|\boldsymbol{k}|}\begin{pmatrix} k_1 \\ k_2 \\ k_3 \end{pmatrix}, \tag{11}$$

The $3\times 1$ column matrices $\varepsilon(\boldsymbol{k},\lambda)$ stand for the three polarization vectors of the electromagnetic field, where $\varepsilon(\boldsymbol{k},0)$ (parallel to $\boldsymbol{k}$) denotes the longitudinal polarization vector, while $\varepsilon(\boldsymbol{k},\pm 1)$ (perpendicular to $\boldsymbol{k}$) correspond to the right- and left-hand circular polarization vectors, respectively. It is easy to prove that $(\boldsymbol{\tau}\cdot\boldsymbol{k}/|\boldsymbol{k}|)\varepsilon(\boldsymbol{k},\lambda) = \lambda\varepsilon(\boldsymbol{k},\lambda)$, and then $\lambda=\pm1,0$ represent the spin projections in the direction of $\boldsymbol{k}$. When the electromagnetic field is described by a 4D electromagnetic potential, there involves four polarization vectors together describing four kinds of photons; while described by the photon wave function (defined in terms of electromagnetic field intensities), there only involves three polarization vectors labeled by the parameters $\lambda=\pm1,0$, where the $\lambda=0$ solution describes the admixture of the longitudinal and scalar photons, and the $\lambda=\pm1$ solutions describe the two transverse photons. Using Eq. (11) one has the orthonormality and completeness relations

$$\begin{cases}\varepsilon^{+}(\boldsymbol{k},\lambda)\varepsilon(\boldsymbol{k},\lambda')=\delta_{\lambda\lambda'}\\ \sum_{\lambda}\varepsilon(\boldsymbol{k},\lambda)\varepsilon^{+}(\boldsymbol{k},\lambda)=I_{3\times3}\end{cases}. \tag{12}$$

Using Eqs. (10) and (12) one has the orthonormality and completeness relations

$$\begin{cases}f^{+}(k,\lambda)f(k,\lambda')=g^{+}(k,\lambda)g(k,\lambda')=\delta_{\lambda\lambda'}\\ f^{+}(k,\lambda)g(k_0,-\boldsymbol{k},\lambda')=g^{+}(k_0,-\boldsymbol{k},\lambda)f(k,\lambda')=0\end{cases}, \tag{13}$$

$$\sum_{\lambda}[f(k,\lambda)f^{+}(k,\lambda)+g(k_0,-\boldsymbol{k},\lambda)g^{+}(k_0,-\boldsymbol{k},\lambda)]=I_{6\times6}, \tag{14}$$

where $I_{6\times6}$ denotes the 6×6 unit matrix, $f(k,\lambda)=f(k_0,\boldsymbol{k},\lambda)$ (and so on). Consider that antiphotons are identical with photons, using Eq. (9a) one can construct the general solution of Eq. (6) as follows:

$$\begin{aligned}\varphi(x)&=\sqrt{1/2}\sum_{\boldsymbol{k},\lambda}[a(k,\lambda)\phi_{k,\lambda}+a^{+}(k,\lambda)\phi_{-k,\lambda}]\\ &=\sum_{\boldsymbol{k},\lambda}\sqrt{\omega/2V}f(k,\lambda)[a(k,\lambda)\exp(-ik\cdot x)+a^{+}(k,\lambda)\exp(ik\cdot x)]\end{aligned}. \tag{15}$$

The commutation relations between $a(k,\lambda)$ and $a^{+}(k',\lambda')$ must be in agreement with the traditional QED, rather than introduced by an additional hypothesis. Let $A^{\mu}(x)$ be a 4D electromagnetic potential satisfying Eq. (6), in the Lorentz gauge condition $\partial_{\mu}A^{\mu}=0$, it can be expanded as

$$A^{\mu}(x)=\sum_{\boldsymbol{k}}\sum_{s=0}^{3}\frac{1}{\sqrt{2\omega V}}e^{\mu}(\boldsymbol{k},s)[c(\boldsymbol{k},s)\exp(-ik\cdot x)+c^{+}(\boldsymbol{k},s)\exp(ik\cdot x)],\tag{16}$$

where $e^{\mu}(\boldsymbol{k},s)$'s are four 4D polarization vectors with the four indices of $s$=0,1,2,3 corresponding to four kinds of photons, respectively. The vector representation of $\varepsilon(\boldsymbol{k},\lambda)$ [given by Eq. (11)] is written as $\boldsymbol{\eta}(\boldsymbol{k},\lambda)$ ($\lambda=\pm1,0$), that is

$$\begin{cases}\boldsymbol{\eta}(\boldsymbol{k},1)=\boldsymbol{\eta}^{*}(\boldsymbol{k},-1)=\dfrac{1}{\sqrt{2}|\boldsymbol{k}|}(\dfrac{k_1k_3-ik_2|\boldsymbol{k}|}{k_1-ik_2},\dfrac{k_2k_3+ik_1|\boldsymbol{k}|}{k_1-ik_2},-(k_1+ik_2))\\ \boldsymbol{\eta}(\boldsymbol{k},0)=\dfrac{\boldsymbol{k}}{|\boldsymbol{k}|}=\dfrac{1}{|\boldsymbol{k}|}(k_1,k_2,k_3)\end{cases}.\tag{17}$$

Using Eq. (17), the four 4D polarization vectors $e^{\mu}(\boldsymbol{k},s)$ ($s=0,1,2,3$) can be chosen as

$$\begin{cases}e^{\mu}(\boldsymbol{k},0)=(1,0,0,0),\ \ e^{\mu}(\boldsymbol{k},1)=(0,\boldsymbol{\eta}(\boldsymbol{k},1))\\ e^{\mu}(\boldsymbol{k},2)=(0,\boldsymbol{\eta}(\boldsymbol{k},-1)),\ \ e^{\mu}(\boldsymbol{k},3)=(0,\boldsymbol{\eta}(\boldsymbol{k},0))\end{cases}.\tag{18}$$

As we know, the electromagnetic field intensities $\boldsymbol{E}$ and $\boldsymbol{B}$ are defined as

$$\boldsymbol{E}=-\nabla A^{0}-\partial_t\boldsymbol{A},\ \ \boldsymbol{B}=\nabla\times\boldsymbol{A}.\tag{19}$$

Applying Eqs. (16)-(19), and note that $\boldsymbol{e}(\boldsymbol{k},0)=(0,0,0)$, $e^{0}(\boldsymbol{k},s)=\delta_{s0}$ ($s=0,1,2,3$), $\boldsymbol{k}=|\boldsymbol{k}|\boldsymbol{\eta}(\boldsymbol{k},0)=\omega\boldsymbol{\eta}(\boldsymbol{k},0)$, and $\lambda\boldsymbol{\eta}(\boldsymbol{k},\lambda)=0$ for $\lambda=0$, one can obtain:

$$\begin{cases}\boldsymbol{E}=\sum_{\boldsymbol{k},\lambda}\sqrt{\dfrac{\omega}{V}}[\dfrac{1}{\sqrt{1+\lambda^{2}}}\boldsymbol{\eta}(\boldsymbol{k},\lambda)][a(\boldsymbol{k},\lambda)\exp(-ik\cdot x)+a^{+}(\boldsymbol{k},\lambda)\exp(ik\cdot x)]\\ \boldsymbol{B}=\sum_{\boldsymbol{k},\lambda}\sqrt{\dfrac{\omega}{V}}[\dfrac{-i}{\sqrt{1+\lambda^{2}}}\lambda\boldsymbol{\eta}(\boldsymbol{k},\lambda)][a(\boldsymbol{k},\lambda)\exp(-ik\cdot x)+a^{+}(\boldsymbol{k},\lambda)\exp(ik\cdot x)]\end{cases},\tag{20}$$

where

$$\begin{cases}a(\boldsymbol{k},1)=ic(\boldsymbol{k},1)\\ a(\boldsymbol{k},-1)=ic(\boldsymbol{k},2)\\ a(\boldsymbol{k},0)=i[c(\boldsymbol{k},3)-c(\boldsymbol{k},0)]/\sqrt{2}\end{cases}.\tag{21}$$

Comparing Eq. (15) with the matrix representation of Eq. (20) (note that the matrix representation of $\boldsymbol{\eta}(\boldsymbol{k},\lambda)$ is $\varepsilon(\boldsymbol{k},\lambda)$), Eqs. (10) and (20) imply $\varphi(x)=\frac{1}{\sqrt{2}}\begin{pmatrix}\boldsymbol{E}\\ i\boldsymbol{B}\end{pmatrix}$, which is in agreement with Eq. (3). Eq. (21) shows that the $\lambda = \pm 1$ solutions describe two kinds of transverse photons ($s$=1, 2), while the $\lambda = 0$ photons correspond to the admixture of the longitudinal ($s$=3) and scalar ($s$=0) photons. According to QED, only those state vectors (say, $|\Phi\rangle$) are admitted for which the expectation value of the Lorentz gauge condition is satisfied: $\langle\Phi|\partial^\mu A_\mu|\Phi\rangle = 0$, which implies that

$$\langle\Phi|a(\boldsymbol{k},0)|\Phi\rangle = i\frac{1}{\sqrt{2}}\langle\Phi|[c(\boldsymbol{k},3)-c(\boldsymbol{k},0)]|\Phi\rangle = 0. \tag{22}$$

For the moment, the transversality condition $\nabla\cdot\boldsymbol{E}=0$ is no longer valid, instead one has $\langle\Phi|\nabla\cdot\boldsymbol{E}|\Phi\rangle = 0$, i.e., $\langle\Phi||\boldsymbol{k}|a(\boldsymbol{k},0)|\Phi\rangle = \langle\Phi|\omega a(\boldsymbol{k},0)|\Phi\rangle = 0$, which is valid for arbitrary $k^\mu = (\omega,\boldsymbol{k})$ provided that Eq. (22) is true. Moreover, Eqs. (8) and (22) imply that

$$\omega_0 = |\boldsymbol{k}_0| = \langle\Phi|\omega a(\boldsymbol{k},0)|\Phi\rangle = 0. \tag{23}$$

Seeing that the contributions from the longitudinal and scalar photons can cancel each other only via Eq. (22), in the following we rewrite $k_\lambda^\mu = (\omega_\lambda,\boldsymbol{k}_\lambda)$ ($\lambda = \pm 1, 0$) as $k^\mu = (\omega,\boldsymbol{k})$ uniformly. According to QED, there have the following commutation relations:

$$[c(\boldsymbol{k},s),c^+(\boldsymbol{k}',s')] = -g_{ss'}\delta_{\boldsymbol{kk}'}, \tag{24}$$

with the others vanishing. Applying Eqs. (21) and (24), one can obtain the following commutation relations

$$[a(k,\lambda),a^+(k',\lambda')] = \delta_{\boldsymbol{kk}'}\delta_{\lambda\lambda'},\ \lambda,\lambda' = \pm 1, \tag{25a}$$

with other commutators vanishing. Likewise, using Eq. (9b) one can provide another general solution for Eq. (6) as follows:

$$\begin{aligned}\varphi'(x) &= \sqrt{1/2}\sum_{\boldsymbol{k},\lambda}[b(k,\lambda)\phi'_{k,\lambda} + b^+(k,\lambda)\phi'_{-k,\lambda}] \\ &= \sum_{\boldsymbol{k},\lambda}\sqrt{\omega/2V}\,g(k,\lambda)[b(k,\lambda)\exp(-ik\cdot x) + b^+(k,\lambda)\exp(ik\cdot x)]\end{aligned}. \tag{26}$$

Similarly, one can obtain the following commutation relations

$$[b(k,\lambda), b^{+}(k',\lambda')] = \delta_{\boldsymbol{kk}'}\delta_{\lambda\lambda'},\ \lambda,\lambda' = \pm 1, \tag{25b}$$

with other commutators vanishing. However, using Eqs. (9a) and (9b) together, one can also choose a set of fundamental solutions as follows:

$$\begin{cases} \phi_{+k,\lambda}(x) = (\omega/V)^{1/2} f(k,\lambda)\exp(-ik\cdot x) \\ \phi'_{-k,\lambda}(x) = (\omega/V)^{1/2} g(k,\lambda)\exp(ik\cdot x) \end{cases},\ \lambda = \pm 1, 0\,. \tag{27}$$

They also form an orthonormal and complete set, that is

$$\begin{cases} \int \phi^{+}_{+k,\lambda}(x)\phi_{+k',\lambda'}(x)d^3x = \int \phi'^{+}_{-k,\lambda}(x)\phi'_{-k',\lambda'}(x)d^3x = \omega\delta_{\lambda\lambda'}\delta_{\boldsymbol{kk}'} \\ \int \phi^{+}_{+k,\lambda}(x)\phi'_{-k'_0,\boldsymbol{k}',\lambda'}(x)d^3x = \int \phi'^{+}_{-k_0,\boldsymbol{k},\lambda}(x)\phi_{+k',\lambda'}(x)d^3x = 0 \end{cases}, \tag{28}$$

$$\sum_{\lambda}\int [\phi_{k,\lambda}(x)\phi^{+}_{k,\lambda}(x) + \phi'_{-k_0,\boldsymbol{k},\lambda}(x)\phi'^{+}_{-k_0,\boldsymbol{k},\lambda}(x)]d^3x = \omega I_{6\times 6}\,, \tag{29}$$

Therefore, the general solution of Eq. (6) can also be expanded in terms of Eq. (27):

$$\begin{aligned} \psi(x) &= \sum_{\boldsymbol{k},\lambda}[a(k,\lambda)\phi_{k,\lambda} + b^{+}(k,\lambda)\phi'_{-k,\lambda}] \\ &= \sum_{\boldsymbol{k},\lambda}\sqrt{\omega/V}[a(k,\lambda)f(k,\lambda)\exp(-ik\cdot x) + b^{+}(k,\lambda)g(k,\lambda)\exp(ik\cdot x)] \end{aligned}\,. \tag{30}$$

For our purpose, we will adopt the general solution given by Eq. (30), which can be regarded as a linear combination of Eqs. (15) and (26). In the present formalism, the symmetry between the positive- and negative-energy solutions correspond to the duality between the electric and magnetic fields, rather than to the usual particle-antiparticle symmetry. Therefore, in Eq. (30), we call the positive-energy components the photon solution, while call the negative-energy components the dual-photon solution, the parity of photons is opposite to that of dual-photons, just as we will show later. If the positive-energy solutions are regarded as the fields produced by electric multipole moments, then the negative-energy solutions can be regarded as the fields produced by magnetic multipole moments, which is due to the fact that there exists a duality between the fields produced by electric multipole moments and those produced by magnetic multipole moments, and the parities of the fields produced by an electric multipole moment are just opposite to those produced by a magnetic multipole moment of the same order [31]. The modes of photons are different from those of dual-photons, then using Eqs. (25a) and (25b) one has

$$[a(k,\lambda),a^{+}(k',\lambda')]=[b(k,\lambda),b^{+}(k',\lambda')]=\delta_{\boldsymbol{kk}'}\delta_{\lambda\lambda'},\ \ \lambda,\lambda'=\pm 1, \tag{31}$$

with other commutators vanishing. In particular, one has $[a(k,0),a^{+}(k',0)]=0$ and $[b(k,0),b^{+}(k',0)]=0$, which is due to the fact that the contributions of the longitudinal and scalar photons cancel each other. Nevertheless, as we know, the longitudinal and scalar photons can exist in the form of *virtual* particles.

It is significant to note that the Dirac-like equation (6) is valid for all source-free electromagnetic fields (e.g., for both time-varying and static fields). For example, at the level of classic field theory, applying $k_0^{\mu}=(0,0,0,0)$ and the Maxwell equations (1) and (2), one can show that the $\lambda=0$ fields (say, $\boldsymbol{E}_{\text{static}}$) satisfy both $\nabla\times\boldsymbol{E}_{\text{static}}=0$ and $\nabla\cdot\boldsymbol{E}_{\text{static}}=0$, which is the characteristic feature of a static field outside a source; at the level of quantum field theory, using Eq. (22) one has $\langle\Phi|\nabla\cdot\boldsymbol{E}_{\text{static}}|\Phi\rangle=\langle\Phi|\nabla\times\boldsymbol{E}_{\text{static}}|\Phi\rangle=0$. In other words, nearing a source, but not including the source, there also contain the $\lambda=0$ photons consisting of *virtual* longitudinal and scalar photons (as we know, the Coulomb interaction arises from the combined exchange of *virtual* longitudinal and scalar photons [32]). Therefore, in our formalism, the $\lambda=\pm 1,0$ solutions of the Dirac-like equation (6) together describe the radiation and static fields outside a source in a unified way.

## IV. SYMMETRY PROPERTIES OF THE DIRAC-LIKE EQUATION

Now, let us introduce a Lorentz scalar as follows (see Appendix A)

$$\pounds = \bar{\psi}(x)(\mathrm{i}\beta^{\mu}\partial_{\mu})\psi(x)\,. \tag{32}$$

We call it the pseudo-Lagrangian density, and call $A = \int \pounds d^4x$ the pseudo-action. Applying the principle of least pseudo-action in $A = \int \pounds d^4x$ one can easily obtain the Dirac-like equation (6), but notice that the dimension of the pseudo-Lagrangian density is $[1/\text{length}]^5$ instead of $[1/\text{length}]^4$, which implies that the pseudo-action has the dimension of $[1/\text{length}]^1$ rather than $[1/\text{length}]^0$. In our formalism, all the symmetries of the Dirac-like Eq. (6) are equivalent to the ones of the pseudo-Lagrangian density given by Eq. (32), and then let us discuss some symmetries of $\pounds$ (its Lorentz invariance is shown in Appendix A).

(1) *Space-time translation symmetry*. It is easy to show that the pseudo-Lagrangian density (32) is invariant under the space-time translations $x^{\mu} \to x'^{\mu} = x^{\mu} + \varepsilon^{\mu}$ ($\varepsilon^{\mu}$'s are infinitesimal real constants). However, related to the fact that the pseudo-Lagrangian density has dimension of $[1/\text{length}]^5$ rather than $[1/\text{length}]^4$, the corresponding conserved current *density* (say, $\Theta^{\mu\nu}$) is not a canonical energy-momentum tensor, and the conserved current $\Xi^{\mu\nu} = \int \Theta^{\mu\nu} d^3x = \int \varpi^{\mu} p^{\nu} d^3x$ is a 4D tensor with the dimension of $[1/\text{length}]^2$, where $p^{\nu}$ is a 4D momentum and $\varpi^{\mu} = ((\boldsymbol{E}^2 + \boldsymbol{B}^2)/2, \boldsymbol{E} \times \boldsymbol{B})$ (not a 4D vector).

(2) *Generalized gauge transformation*. As we know, under the usual gauge transformation, the electromagnetic field intensities $\boldsymbol{E}$ and $\boldsymbol{B}$ (and thus the photon wave function $\psi(x)$) are invariant, such that the pseudo-Lagrangian density is also invariant. On the other hand, it is easy to prove that, under the transformation

$$\psi(x) \to \exp(-i\theta)\psi(x)\,, \tag{33}$$

the pseudo-Lagrangian density (32) [and then the Dirac-like Eq. (6)] is invariant, where $\theta$ is a real constant. The transformation (33) applied to the photon wave function can be called as *generalized gauge transformation*, because it is formally similar to the usual gauge transformation applied to the Dirac field. Under the generalized gauge transformation (33), one can easily show that the corresponding conserved current $J_\mu$ (rather than the current *density*) is a 4D vector:

$$J^\mu = \int \bar{\psi}(x)\beta^\mu\psi(x)d^3x = \int ((\boldsymbol{E}^2 + \boldsymbol{B}^2)/2, \boldsymbol{E}\times\boldsymbol{B})d^3x\,, \tag{34}$$

Therefore, the conserved current related to the generalized gauge symmetry of Eq. (32) is the 4D momentum of the electromagnetic field. In the post-Newtonian theory of gravity, the energy plays the role of gravitation charge, and then one might take the post-Newtonian gravity field as the generalized gauge field transferring the gravitational interactions between photons. Seeing that the photon field itself is the usual gauge field transferring the electromagnetic interactions between the electrons, the generalized gauge field is the gauge field of the usual gauge field. In our source-free case, $\boldsymbol{g} = \boldsymbol{E}\times\boldsymbol{B}$ and $h = (\boldsymbol{E}^2 + \boldsymbol{B}^2)/2$ satisfy the differential form of Poynting's theorem, i.e., $\nabla\cdot\boldsymbol{g} + \partial_t h = 0$, it is not Lorentz covariant because $(h, \boldsymbol{g})$ does not form a 4D vector. On the other hand, the 4D current vector $J^\mu = \int (h, \mathbf{g})d^3x$ given by Eq. (34) satisfies the integral form of Poynting's theorem, i.e., $\partial_\mu J^\mu = 0$, it is Lorentz covariant, then we regard $\partial_\mu J^\mu = 0$ as the continuity equation of photons in our formalism.

(3) *Space inversion symmetry*. Under the parity transformation of $\boldsymbol{x}\to-\boldsymbol{x}$ (say, P), one can show that Eq. (32) [and then the Dirac-like equation (6)] is invariant, where the field quantity $\psi(x)$ transforms as

$$\mathrm{P}\psi(\boldsymbol{x},t)\mathrm{P}^{-1} = \beta^0\psi(-\boldsymbol{x},t)\,, \tag{35}$$

where $\mathrm{P}^{-1}$ represents the inverse of P. Inserting the expansion (30) into Eq. (35), one can obtain ( $\lambda = \pm 1, 0$ )

$$\mathrm{P}a(k_0,\boldsymbol{k},\lambda)\mathrm{P}^{-1} = a(k_0,-\boldsymbol{k},\lambda)\,,\ \ \mathrm{P}b(k_0,\boldsymbol{k},\lambda)\mathrm{P}^{-1} = -b(k_0,-\boldsymbol{k},\lambda)\,. \tag{36}$$

As mentioned before, under the duality transformation of exchanging the roles of $\boldsymbol{E}$ and $\boldsymbol{B}$, one can obtain the electric multipole fields from the magnetic multipole fields (up to a multiplying constant at most), and vice versa. In addition, in our case the symmetry between the positive- and negative-energy solutions corresponds to the duality between the electric and magnetic fields, rather than to the usual particle-antiparticle symmetry. Therefore, Eq. (36) is exactly in agreement with the traditional conclusions that, the parities of the fields produced by an electric multipole moment are just opposite to the parities of the fields produced by a magnetic multipole moment of the same order [31].

(4) *Time reversal symmetry*. Under the time reversal transformation of $t \to -t$ (denoted by $\mathrm{T}$), one can show that Eq. (32) [and then the Dirac-like equation (6)] is invariant, and the field quantity $\psi(x)$ transforms as

$$\mathrm{T}\psi(\boldsymbol{x},t)\mathrm{T}^{-1} = \psi(\boldsymbol{x},-t)\,, \tag{37}$$

it implies that

$$\begin{cases} \mathrm{T}a(k_0,\boldsymbol{k},\pm 1)\mathrm{T}^{-1} = a(k_0,-\boldsymbol{k},\mp 1), \mathrm{T}a(k,0)\mathrm{T}^{-1} = a(k,0) \\ \mathrm{T}b(k_0,\boldsymbol{k},\pm 1)\mathrm{T}^{-1} = b(k_0,-\boldsymbol{k},\mp 1), \mathrm{T}b(k,0)\mathrm{T}^{-1} = b(k,0) \end{cases}. \tag{38}$$

(5) *Dual transformation*. It is easy to show that, both the dual transformation of $\boldsymbol{E} \leftrightarrow i\boldsymbol{B}$ and that of $\boldsymbol{E} \leftrightarrow -i\boldsymbol{B}$ correspond to the transformation

$$\psi \to \psi' = \mathrm{C}\psi\mathrm{C}^{-1} \equiv \begin{pmatrix} 0 & I_{3\times 3} \\ -I_{3\times 3} & 0 \end{pmatrix}\psi^*\,. \tag{39}$$

Eq. (32) [and then the Dirac-like equation (6)] is invariant under the transformation (39). For example, under the transformation of $\boldsymbol{E} \leftrightarrow -i\boldsymbol{B}$, one has

$$\begin{cases} \mathrm{C}a(k,\lambda)\mathrm{C}^{-1} = b(k,-\lambda) \\ \mathrm{C}b(k,\lambda)\mathrm{C}^{-1} = -a(k,-\lambda) \end{cases}. \tag{40}$$

In our formalism, the symmetry between the positive- and negative-energy solutions corresponds to the duality between the electric and magnetic fields, and then the dual transformation is related to the photon field in the same way that the charge-conjugation transformation is related to the Dirac field.

## V. FIELD QUANTIZATION AND ZB OF PHOTONS

In our formalism, the 4D momentum of the electromagnetic field, as the conserved current related to the generalized gauge transformation (33), is given by Eq. (34). Substituting Eq. (30) into Eq. (34), and applying Eq. (31) and Eq. (22) [it is also valid for $b(k,0)$], one has

$$J^0 = \sum_{\boldsymbol{k},\lambda=\pm 1} \omega[a^+(k,\lambda)a(k,\lambda) + b^+(k,\lambda)b(k,\lambda) + 1], \tag{41}$$

Here, $\hat{N}_e(k,\lambda) = a^+(k,\lambda)a(k,\lambda)$ and $\hat{N}_m(k,\lambda) = b^+(k,\lambda)b(k,\lambda)$ represent the number operator of photons excited by an electric multipole moment and that of photons excited by a magnetic multipole moment of the same order, respectively.

We study the ZB of photons via the 3D momentum of the electromagnetic field, such that the concept of position operator of photons is not necessary. Using Eq. (34) one has

$$\begin{aligned} \boldsymbol{J} &= \int \psi^+(x)\beta_0\boldsymbol{\beta}\psi(x)d^3x \\ &= \sum_{\boldsymbol{k},\lambda,\lambda'} \omega[a^+(k,\lambda)a(k,\lambda') + b(k,\lambda)b^+(k,\lambda')]\boldsymbol{R}(\lambda,\lambda') \\ &+ \sum_{\boldsymbol{k},\lambda,\lambda'} \omega[b(k,\lambda)a(k_0,-\boldsymbol{k},\lambda')\exp(-2i\omega t)]\boldsymbol{T}(\lambda,\lambda') \\ &+ \sum_{\boldsymbol{k},\lambda,\lambda'} \omega[a^+(k,\lambda)b^+(k_0,-\boldsymbol{k},\lambda')\exp(2i\omega t)]\boldsymbol{T}(\lambda,\lambda') \end{aligned}, \tag{42}$$

where $\boldsymbol{R}(\lambda,\lambda')$ and $\boldsymbol{T}(\lambda,\lambda')$ are two 3D vectors, they are

$$\begin{cases} \boldsymbol{R}(\lambda,\lambda') = \dfrac{\lambda+\lambda'}{\sqrt{(1+\lambda^2)(1+\lambda'^2)}}\varepsilon^+(\boldsymbol{k},\lambda)\boldsymbol{\tau}\varepsilon(\boldsymbol{k},\lambda') \\ \boldsymbol{T}(\lambda,\lambda') = \dfrac{(1+\lambda\lambda')}{\sqrt{(1+\lambda^2)(1+\lambda'^2)}}\varepsilon^+(\boldsymbol{k},\lambda)\boldsymbol{\tau}\varepsilon(-\boldsymbol{k},\lambda') \end{cases}, \quad \lambda,\lambda' = \pm1, 0\,. \tag{43}$$

Using Eqs. (4) and (11), one has

$$\begin{cases} \boldsymbol{R}(1,-1) = \boldsymbol{R}(-1,1) = \boldsymbol{R}(0,0) = (0,0,0),\ \boldsymbol{R}(1,1) = \boldsymbol{R}(-1,-1) = \boldsymbol{k}/\omega \\ \boldsymbol{R}(0,1) = \boldsymbol{R}(-1,0) = \boldsymbol{R}^*(1,0) = \boldsymbol{R}^*(0,-1) = -\boldsymbol{\eta}(\boldsymbol{k},1)/\sqrt{2} \\ \boldsymbol{T}(1,-1) = \boldsymbol{T}(-1,1) = \boldsymbol{T}(0,0) = \boldsymbol{T}(1,1) = \boldsymbol{T}(-1,-1) = (0,0,0) \\ \boldsymbol{T}(-1,0) = -\boldsymbol{T}^*(1,0) = -\boldsymbol{\eta}(\boldsymbol{k},1)/\sqrt{2}\,, \boldsymbol{T}(0,1) = -\boldsymbol{T}^*(0,-1) = \boldsymbol{\eta}(-\boldsymbol{k},1)/\sqrt{2} \end{cases}, \tag{44}$$

Substituting Eqs. (44) and (43) into (42), one has

$$\begin{aligned} \boldsymbol{J} &= \sum_{\boldsymbol{k},\lambda=\pm1} \boldsymbol{k}[a^+(k,\lambda)a(k,\lambda) + b^+(k,\lambda)b(k,\lambda)] \\ &- \sum_{\boldsymbol{k},\lambda=\pm1} \{\frac{\omega}{\sqrt{2}}\boldsymbol{\eta}(\boldsymbol{k},\lambda)[a^+(k,0)a(k,\lambda) + b(k,0)b^+(k,\lambda)] + \text{H.c.}\} \\ &+ \sum_{\boldsymbol{k},\lambda=\pm1} \frac{\lambda\omega}{\sqrt{2}}\boldsymbol{\eta}(\boldsymbol{k},-\lambda)[b(k,\lambda)a(k_0,-\boldsymbol{k},0)\exp(-2i\omega t) + a^+(k,\lambda)b^+(k_0,-\boldsymbol{k},0)\exp(2i\omega t)] \\ &+ \sum_{\boldsymbol{k},\lambda=\pm1} \frac{\lambda\omega}{\sqrt{2}}\boldsymbol{\eta}(-\boldsymbol{k},\lambda)[b(k,0)a(k_0,-\boldsymbol{k},\lambda)\exp(-2i\omega t) + a^+(k,0)b^+(k_0,-\boldsymbol{k},\lambda)\exp(2i\omega t)] \end{aligned}, \tag{45}$$

where H.c. stands for the Hermitian conjugate of the previous term. Note that $\boldsymbol{\eta}(\boldsymbol{k},\lambda)$'s are the vector representations of the 3×1 column matrices $\varepsilon(\boldsymbol{k},\lambda)$ given by Eq. (17) (and so on), they stand for the three polarization vectors of the electromagnetic field, where $\lambda = \pm1, 0$ represent the spin projections in the direction of $\boldsymbol{k}$. The third and fourth terms on the right-hand sides of Eq. (45) are perpendicular to $\boldsymbol{k}$ and represent the transverse ZB motion of photons. As mentioned before, the $\lambda = 0$ photon corresponds to the admixture of the longitudinal and scalar photons, and it can exist in the form of virtual photons. Obviously, Eq. (21) [it is also valid for $b(k,0)$] and Eq. (45) together show that there does not exist any ZB motion of photons when not taking into account the influences of virtual photons. On the contrary, when the Coulomb interaction arising from the combined exchange of *virtual* longitudinal and scalar photons is considered, the ZB motions of photons occur. Therefore, according to Eq. (45), the ZB phenomenon of photons stems from the presence of virtual longitudinal and scalar photons. It is interesting to note that, the ZB motions of the Dirac electron are also described by the polarization vectors of the electromagnetic field [26] [in Ref. [26], $\boldsymbol{\eta}(\boldsymbol{k},\pm 1)$ and $\boldsymbol{\eta}(\boldsymbol{k},0)$ are written as $\boldsymbol{\eta}_{\pm}$ and $\boldsymbol{\eta}_{\parallel}$, respectively], and similarly, the ZB of an electron arises from the influence of virtual electron-positron pairs on the electron [26, 33, 34].

Likewise, in the traditional QED, in terms of our notions, the 4D momentum of the electromagnetic field can be expressed as

$$J_{\varphi}^{\mu}=(J_{\varphi}^{0},\boldsymbol{J}_{\varphi})=\int\varphi^{+}(x)\beta_{0}\beta^{\mu}\varphi(x)d^{3}x\,, \tag{46}$$

where $\varphi(x)$ is given by Eq. (15). Substituting Eq. (15) into Eq. (46) one has

$$\begin{aligned}\boldsymbol{J}_{\varphi}=&\sum_{\boldsymbol{k},\lambda=\pm1}\boldsymbol{k}a^{+}(k,\lambda)a(k,\lambda)-\sum_{\boldsymbol{k},\lambda=\pm1}\frac{\omega}{\sqrt{2}}\boldsymbol{\eta}(\boldsymbol{k},\lambda)[a^{+}(k,0)a(k,\lambda)+\text{H.c.}]\\&-\sum_{\boldsymbol{k},\lambda=\pm1}\frac{\lambda\omega}{2\sqrt{2}}\boldsymbol{\eta}(\boldsymbol{k},-\lambda)[a(k,\lambda)a(k_{0},-\boldsymbol{k},0)\exp(-2i\omega t)+\text{H.c.}]\\&+\sum_{\boldsymbol{k},\lambda=\pm1}\frac{\lambda\omega}{2\sqrt{2}}\boldsymbol{\eta}(-\boldsymbol{k},\lambda)[a(k,0)a(k_{0},-\boldsymbol{k},\lambda)\exp(-2i\omega t)+\text{H.c.}]\end{aligned}\,, \tag{47}$$

Therefore, in the traditional formalism, the ZB motions of photons also occur in the presence of virtual longitudinal and scalar photons.

## VI. CONCLUSIONS AND DISCUSSIONS

Making use of a photon wave function with the meaning of energy-density amplitude as well as its Lorentz-covariant equation of motion (i.e., the Dirac-like equation), one can provide an unified and complete description for all kinds of electromagnetic fields outside a source, including the time-varying and static fields, the transverse and longitudinal fields, and those generated by an electrical and magnetic multipole moments, etc. In our formalism, the symmetry between the positive- and negative- energy solutions corresponds to the duality between the electric and magnetic fields, rather than to the usual particle-antiparticle symmetry, and the photon wave function corresponds to the $(1,0)\oplus(0,1)$ representation of the Lorentz group (instead of the chiral representation presented in the previous literatures). Some symmetry properties of our theory are discussed from a new point of view.

On the one hand, our results are obtained starting from fundamental principles and without resorting to any additional hypothesis, and then they are natural; on the other hand, Eqs. (45) and (47) show that, only for those real photons in the presence of *virtual* longitudinal and scalar photons, their ZB motions can occur. Therefore the photonic ZB may occur in the following two cases:

1). As we know, the Coulomb interaction arises from the combined exchange of *virtual* longitudinal and scalar photons [32], then for a real photon passing through a strong Coulomb field, its ZB motion may occur, which may be due to some nonlinear optical effects.

2). In our formalism, the conserved current related to the *generalized* gauge transformation (33) is the 4D momentum of the electromagnetic field, and then it plays the role of gravitation charge. Taking the post-Newtonian gravity field as the generalized gauge field transferring the gravitational interactions between photons, one can present an possible example for the ZB motions of photons as follows: around a (real) transverse photon, virtual photon pairs are continuously created and annihilated in a gravity field, the transverse photon can annihilate with the longitudinal (or scalar) photon of a virtual pair consisting of a transverse photon and a longitudinal (or scalar) photon, while the transverse photon of the virtual pair which is left over now replaces the original transverse photon, by such an exchange the ZB of the real transverse photon occurs. Therefore, for photons in the vicinity of a black hole, their ZB can occur, provided that there is such a physical process: a gravition becomes a (virtual) photon pair, and then the photon pair annihilates and becomes a gravition. In the present case, the ZB of a photon is due to the influence of virtual photon pairs (arising from the gravitational vacuum fluctuations) on the photon. As an analogy, as we know, the ZB of an electron is due to the influence of virtual electron-positron pairs (arising from the electromagnetic vacuum fluctuations) on the electron [26], here the electromagnetic field is related to the electron in the similar way that the gravitational field is related to the photon.

Eqs. (45) and (47) and Ref. [26] show that, there are some common properties between the ZB motions of a photon and those of a Dirac electron. For example, they are both related to the polarization vectors of the electromagnetic field, they are both due to the presence of virtual particles, and their oscillation frequencies are both $2E = 2\omega$ ($\hbar = c = 1$, $E$ denotes the relativistic energy of an electron or a photon). Moreover, to estimate the ZB amplitude of an electron or a photon, for convenience let us consider a monochromatic component at the quantum-mechanical level (for the moment $\boldsymbol{J}/\omega$ can be regarded as the velocity vector of photons), one can show that both the ZB amplitude of an electron and that of a photon are of the order of $1/2E = 1/2\omega$. In contrast to the photonic ZB, for an electron one has $2E \geq 2m$ ($m$ denotes the rest mass of the electron), and then the electron's ZB has a very small amplitude and a very high frequency. However, this does not imply that the measurement of photonic ZB is easier than that of the electron's ZB, because the motion velocity of an electron is usually far smaller than the one of photons, and only in the presence of virtual longitudinal and scalar photons, the ZB motion of photons can occur.

In our next work, we will study how to provide an experimental examination for our theoretical results. For example, we will study whether the photonic ZB can be detected by applying photonic crystals. In Refs. [35], the ZB of the Dirac electron is simulated by means of photons near the Dirac point of a two-dimensional (2D) photonic crystal, where the photons satisfy a (2+1)D Dirac equation rather than our Dirac-like equation in a (2+1)D space, and then it might have no relation to the photonic ZB studied in our paper.

The first author (Z. Y. Wang) would like to thank Dr. Changhai Lu, Prof. Erasmo Recami and Prof. Xiangdong Zhang for their helpful discussions. This work was supported by the National Natural Science Foundation of China (Grant No. 60671030).

## APPENDIX A: LORENTZ INVARIANCE OF THE PSEUDO-LAGRANGIAN DENSITY

Under an infinitesimal Lorentz transformation

$$x^{\mu}\to x'^{\mu}=a^{\mu\nu}x_{\nu}\,,\ \ \partial_{\mu}\to\partial'_{\mu}=a_{\mu\nu}\partial^{\nu}\,, \tag{A1}$$

where $a_{\mu\nu}=g_{\mu\nu}+\varepsilon_{\mu\nu}$ with $\varepsilon_{\mu\nu}$ being an real infinitesimal antisymmetric tensor, the photon wave function $\psi(x)$ transforms linearly in the way

$$\psi(x)\to\psi'(x')=\Lambda\psi(x)\,,\ \ \bar{\psi}(x)\to\bar{\psi}'(x')=\bar{\psi}(x)\beta^{0}\Lambda^{+}\beta^{0}\,, \tag{A2}$$

where $\Lambda=1-i\varepsilon^{\mu\nu}\Sigma_{\mu\nu}/2$ with $\Sigma_{\mu\nu}$ being the infinitesimal generators of Lorentz group (in the $(1,0)\oplus(0,1)$ representation). Let $\varepsilon_{lmn}$ denote the full antisymmetric tensor with $\varepsilon_{123}=1$, one has $\Sigma_{lm}=\varepsilon_{lmn}S_{n}$ and $\Sigma_{l0}=-\Sigma_{0l}=i\chi_{l}$ ($l,m,n=1,2,3$), where $\boldsymbol{S}=I_{2\times 2}\otimes\boldsymbol{\tau}$ $[=(S_1,S_2,S_3)]$ is the spin matrix of photons, and $\boldsymbol{\chi}=\beta_0\boldsymbol{\beta}=(\chi_1,\chi_2,\chi_3)$ plays the role of the infinitesimal generators of Lorentz boost. Under the transformations (A1) and (A2), the pseudo-Lagrangian density given by Eq. (32) transforms as

$$\mathcal{L}=\bar{\psi}(x)(i\beta^{\mu}\partial_{\mu})\psi(x)\to\mathcal{L}'=\bar{\psi}(x)\beta^0\Lambda^{+}\beta^0(i\beta^{\mu}a_{\mu\nu}\partial^{\nu})\Lambda\psi(x)\,, \tag{A3}$$

let $\delta\mathcal{L}\equiv\mathcal{L}'-\mathcal{L}$, on can obtain (where $\rho,\lambda=0,1,2,3$)

$$\delta\mathcal{L}(x)=\frac{i}{2}\varepsilon^{\rho\lambda}\bar{\psi}(x)[(\beta_{\rho}\partial_{\lambda}-\beta_{\lambda}\partial_{\rho})-i[\beta^{\mu},\Sigma_{\rho\lambda}]\partial_{\mu}]\psi(x)\,. \tag{A4}$$

A necessary and sufficient condition of $\mathcal{L}$ being Lorentz invariant is $\delta\mathcal{L}(x)=0$. Associating Eq. (A4) with $\delta\mathcal{L}(x)=0$ one has

$$\bar{\psi}(x)(\beta_{\rho}\partial_{\lambda}-\beta_{\lambda}\partial_{\rho})\psi(x)=\bar{\psi}(x)(\mathrm{i}[\beta^{\mu},\Sigma_{\rho\lambda}]\partial_{\mu})\psi(x)\,. \tag{A5}$$

To prove $\delta\mathcal{L}(x)=0$ or Eq. (A5), we show that:

(1) As $\rho=\lambda$, $\varepsilon^{\rho\lambda}=0$, then $\delta\mathcal{L}(x)=0$ is valid;

(2) As $\rho=l$, $\lambda=m$ $(l,m=1,2,3)$, using Eqs. (3) and (4), $\Sigma_{lm}=\varepsilon_{lmn}I_{2\times2}\otimes\tau_n$ as well as $[\tau_l,\tau_m]=i\varepsilon_{lmn}\tau_n$, one has $[\beta^0,\Sigma_{lm}]=0$ and $i[\beta^{\mu},\Sigma_{lm}]\partial_{\mu}=(\beta_l\partial_m-\beta_m\partial_l)$, thus Eq. (A5) is true;

(3) As $\rho=l=1,2,3$, $\lambda=0$ (or alternatively, $\rho=0$, $\lambda=l=1,2,3$), consider that $\Sigma_{l0}=-\Sigma_{0l}=\mathrm{i}\chi_l$ and $\mathrm{i}\beta^{\mu}\partial_{\mu}\psi(x)=0$, Eq. (A5) becomes

$$\bar{\psi}(x)[-(\boldsymbol{\beta}\cdot\partial)\chi_l+\beta^0\partial^l]\psi(x)=0\,. \tag{A6}$$

Using Eqs. (3)-(5) and the transversality condition $\partial^l E_l=\partial^l B_l=0$, it is easy to show that Eq. (A6) is true.

The statements (1)-(3) exhaust all cases, therefore, the pseudo-Lagrangian density with the dimension of $[1/\text{length}]^5$ is Lorentz invariant. In fact, one can also directly prove that Eq. (6) is Lorentz covariant by applying Eqs. (1) and (2).

## APPENDIX B: ANALOGY BETWEEN THE DIRAC SPINOR AND THE PHOTON WAVE FUNCTION

A formal identity between the Dirac equation in the limit of zero mass and Maxwell's equations has been shown in the Ref. [36]. Here, we will present a more detailed analogy between the Dirac spinor and our photon wave function. In relativistic quantum mechanics, a free electron with mass $m$ is described by the Dirac equation for spin-1/2 particles ($\mu,\nu=0,1,2,3$):

$$(\mathrm{i}\gamma^{\mu}\partial_{\mu}-m)\xi(x)=0\,, \tag{B1}$$

where $\gamma^{\mu}$'s are the 4×4 Dirac matrices satisfying the algebra $\gamma^{\mu}\gamma^{\nu}+\gamma^{\nu}\gamma^{\mu}=2g^{\mu\nu}$. Let the four-component spinor $\xi$ be decomposed into two two-component spinors $\chi$ and $\zeta$: $\xi=\begin{pmatrix}\chi\\ \zeta\end{pmatrix}$, in terms of the Pauli matrix vector $\boldsymbol{\sigma}=(\sigma_1,\sigma_2,\sigma_3)$, the Eq.(B1) can also be rewritten as the Maxwell-like form:

$$\begin{cases}(\boldsymbol{\sigma}\cdot\nabla)\chi=(-\partial_t+im)\zeta\\ (\boldsymbol{\sigma}\cdot\nabla)\zeta=(-\partial_t-im)\chi\end{cases}. \tag{B2}$$

Likewise, using Eqs. (3)-(5), the Maxwell equations can be rewritten as the Dirac-like equation (6), i.e.,

$$i\beta^{\mu}\partial_{\mu}\psi(x)=0\,. \tag{B3}$$

The Maxwell equations can also be expressed as the matrix form:

$$\begin{cases}(\boldsymbol{\tau}\cdot\nabla)\boldsymbol{B}=i\partial_t\boldsymbol{E}\\ (\boldsymbol{\tau}\cdot\nabla)\boldsymbol{E}=-i\partial_t\boldsymbol{B}\end{cases}. \tag{B4}$$

Therefore, the Maxwell equations for the free electromagnetic field can be rewritten as the Dirac-like equation; conversely, the Dirac equation for the free electron can be rewritten as the Maxwell-like equations. Comparing the Dirac equation Eq.(B1), with the like-Dirac equation, Eq.(B3), or comparing the Maxwell-like equation, Eq.(B2), with the Maxwell equation, Eq.(B4), one can see that, $\chi$ is related to $\zeta$ in the similar way that $\boldsymbol{E}$ is related to $\boldsymbol{B}$. More concretely, we have the following analogies:

(1) In the Dirac field $\xi = \begin{pmatrix} \chi \\ \zeta \end{pmatrix}$, $\chi$ and $\zeta$ are two different types of 2-component spinor, they respectively correspond to the nonequivalent representations (1/2, 0) and (0, 1/2) of the Lorentz group, and the Dirac equation presents a relation between the two spinors. If we consider parity, then it is no longer sufficient to consider the 2-component spinors $\chi$ and $\zeta$ separately, but the 4-component spinor $\xi = \begin{pmatrix} \chi \\ \zeta \end{pmatrix}$. This 4-component spinor is an irreducible representation of the Lorentz group extended by parity (i.e., the (1/2, 0)$\oplus$(0, 1/2) representation). Likewise, as for the photon field $\psi = \frac{1}{\sqrt{2}}\begin{pmatrix} \boldsymbol{E} \\ i\boldsymbol{B} \end{pmatrix}$, the column matrices $\boldsymbol{E}$ and $i\boldsymbol{B}$ are two different types of three-component spinor, they respectively correspond to the nonequivalent representations (1, 0) and (0, 1) of the Lorentz group, and the Dirac-like equation presents a relation between the two spinors. If we consider parity, then it is no longer sufficient to consider the 3-component spinors $\boldsymbol{E}$ and $i\boldsymbol{B}$ separately, but the 6-component spinor $\psi = \frac{1}{\sqrt{2}}\begin{pmatrix} \boldsymbol{E} \\ i\boldsymbol{B} \end{pmatrix}$. This 6-component spinor is an irreducible representation of the Lorentz group extended by parity (i.e., the (1, 0)$\oplus$(0, 1) representation).

(2) Equation (B4) shows that, a moving or time-varying electric field $\boldsymbol{E}$ induces the magnetic field $\boldsymbol{B}$ and vice versa. Likewise, Eq.(B2) shows that, a moving or time-varying component $\zeta$ induces the component $\chi$ and vice versa.

(3) Under the exchange of $\boldsymbol{E} \leftrightarrow i\boldsymbol{B}$, the free electromagnetic fields have the electricity-

magnetism duality. Under the exchange of $\chi \leftrightarrow \zeta$, the free Dirac fields have particle-antiparticle symmetry.

(4) Both $\psi^{+}\beta^{0}\psi$ $(=\boldsymbol{E}^{+}\boldsymbol{E}-\boldsymbol{B}^{+}\boldsymbol{B})$ and $\xi^{+}\gamma^{0}\xi^{+}(=\zeta^{+}\zeta-\chi^{+}\chi)$ are Lorentz scalars; while both $\psi^{+}\psi$ $(=\boldsymbol{E}^{+}\boldsymbol{E}+\boldsymbol{B}^{+}\boldsymbol{B})$ and $\xi^{+}\xi$ $(=\zeta^{+}\zeta+\chi^{+}\chi)$ are proportional to the number densities.

(5) For the fields produced by an electric source, the electric field $\boldsymbol{E}$ is the large component and the magnetic field $\boldsymbol{B}$ the small component, while for the fields produced by a magnetic source, the magnetic field $\boldsymbol{B}$ is the large component and the electric field $\boldsymbol{E}$ the small component. Similarly, in the electron field, $\zeta$ is the large component and $\chi$ the small component, while in the positron field, $\chi$ is the large component and $\zeta$ the small component.

Moreover, when the wave packet of an electron (or of a positron) is moving with high speeds or varies rapidly, or its size is sufficiently small, or in the present of a strong electromagnetic field, its small component and the corresponding effects cannot be ignored. In particular, an electron wavepacket does contain a positron component, and vice versa, just as that a moving or time-varying electric field is always accompanied by a magnetic field component, and vice versa.

---